\documentclass[a4paper]{jpconf}
\usepackage{graphicx}
\begin{document}
\title{Meson photoproduction from a transversely polarised target at MAMI}

\author{P. Hall Barrientos}

\address{JCMB, University of Edinburgh, Edinburgh, EH9 3JZ, United Kingdom}

\ead{p.hb@ed.ac.uk}

\begin{abstract}
A new transversely polarised frozen spin target has been developed at
MAMI which will be used in conjunction with the photon tagger and the
Crystal Ball detector. The new target permits a major new programme of
accurate measurement of polarisation observables in meson
photoproduction. This contribution presents some of the preliminary
analysis from experimental data taken near the threshold and in the
region of the $\Delta(1232)$ resonance. Observables will be obtained
for the $p\pi^0$ and $n\pi^+$ final states.
\end{abstract}

\section{Introduction}

The dynamics and interactions of the nucleon's constituents are poorly established leading to a global effort aimed at studying it's properties. Although Quantum Chromodynamics (QCD) describes the strong interaction between quarks, it's non-perturbative nature at lower energies results in the necessity to produce theoretical and phenomenological models to describe properties at the nucleon level. 

Understanding the nature of the nucleon's constituents can be achieved by studying their spectrum of excited states, or baryon resonances, manifested from the composite structure of the nucleon. Meson photoproduction is a clean method to access these resonant states. A high energy beam interacts with a nucleon target, excites the nucleon and decays via a meson. Since the excited state of a nucleon is short lived the meson is measured. These measurements exploit polarisation degrees of freedom of the incident photon along with spin degrees of freedom in the target and recoiling nucleons.

Meson photoproduction can be described by 16 experimental observables\cite{Barker}:  the differential cross section, 3 single polarisation observables corresponding to measurements of the beam, target or recoil polarisation and 12 double polarisation observables corresponding to simultaneous measurement of beam-target, beam-recoil or target-recoil polarisations. 

The availability of a polarised photon beam and both longitudinal and transversely polarised targets at Mainz Microtron (MAMI), in Germany, permits accurate measurement of many new single and double polarisation observables in meson photoproduction. These efforts will better establish the excitation spectrum of the nucleon, providing a more complete data set to constrain partial wave analyses of meson photoproduction. Such studies require data over a wide range of photon energies. 

Alongside these endeavours the new facilities make it possible to obtain accurate measurements of single and double polarisation observables near to the reaction threshold and in the region of the well separated lowest lying resonance, the $\Delta(1232)$, to test chiral perturbative QCD theories \cite {chpt} and aspects of QCD such as isospin violation \cite {Aron} \cite {Aronn}.

There is a currently a concerted effort in hadronic physics at most of the major electromagnetic beam facilities to perform a complete or close to complete measurement for various reactions of interest. This requires measurement of at least 8 observables \cite {Dey}. However for meson photoproduction measurements in the threshold region below the 2-pion production threshold, the number of required measurements is predicted to be significantly reduced and the comprehensive new generation of accurate measurements in this region allows precision tests of QCD and QCD based symmetries. The ongoing analysis presented here concerns the measurement of the double polarisation observable F and single polarisation observable T in single pion photoproduction.

\section{Experiment}

In January and February 2010 data was collected at MAMI.The microtron consists of three accelerator stages with the ability to accelerate electrons up to 1.5 GeV \cite{Arends}, for these experimental runs, an electron beam of 150-450 MeV was used. Electrons enter the A2 experimental hall and impinge onto a thin metal radiator. Deceleration of the electrons in the field of the nuclei creates Bremmstrahlung photons. The recoiling electrons are momentum analysed by a magnetic spectrometer, the Glasgow Photon Tagger \cite{Kellie}. Determining the momentum of the recoil electron allows the photon energy to be tagged. The photons are directed onto the polarised butanol (frozen spin) target, which is positioned in the centre of the Crystal Ball (CB) \cite{Peck}, the main detector. 

Reaction products from the interaction of the target and photon are detected by the CB, the Particle Identification Detector (PID) and the TAPS detector array \cite{Novo}. The PID, which directly surrounds the target and detects charged particles, consists of 24 plastic scintillator strips readout at the the upstream end by a PMT. The Crystal Ball is composed of 672 NaI detector elements each shaped of form a truncated pyramid. It covers the polar angular range 20 to 160 degrees. TAPS comprises 385 hexagonal BaF2 elements which cover the forward polar angular region (1 to 20 degrees). The combined system cover over 93$\%$ of $4\pi$ acceptance. A schematic of the combined detector system can be seen in figure 1.

    \begin{figure}[!h]
    	\begin{center}
\includegraphics[height=.37\textheight]{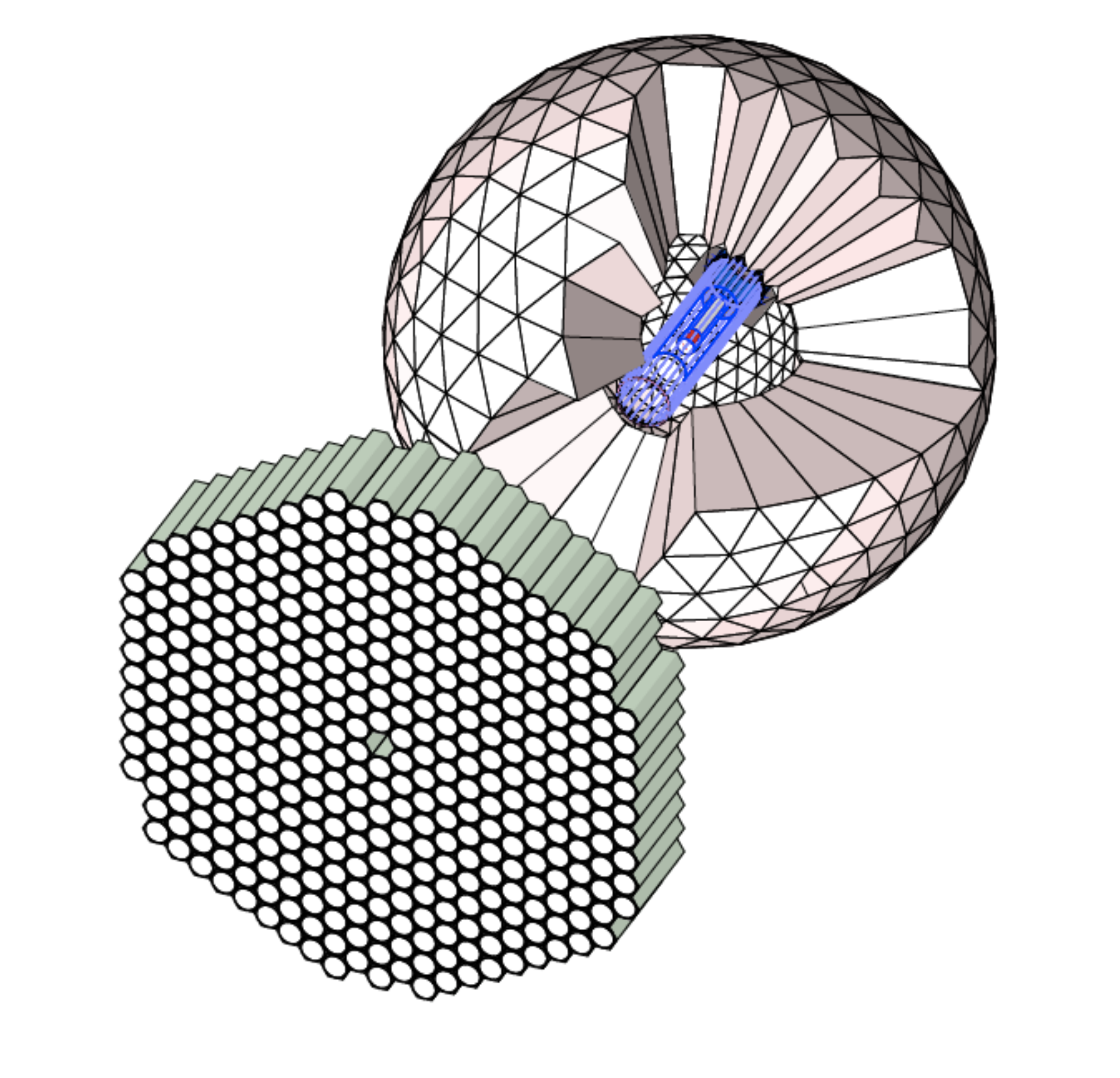}
	\end{center}
    \caption{A schematic representation of the detector system showing TAPS and the Crystal Ball with the PID inside.}
      \end{figure}

The target material in the polarised target consists of beads of butanol doped with a free radical. The target is polarised in a high field of 5 T at a low temperature of ($< $1 K),  polarisation of the free radicals are transferred to the protons using microwave radiation in a process known as Dynamic Nuclear Polarisation \cite{Crabb}. Once polarised the target is further cooled to 27 mK and held in a polarisation mode by a superconducting niobium coil, producing a highly uniform holding field of around 0.6 T. The polarisation of the target is measured by the technique of nuclear magnetic resonance (NMR) via NMR coils, located in the target assembly.

    \begin{figure}[!h]
    	\begin{center}
\includegraphics[height=.37\textheight]{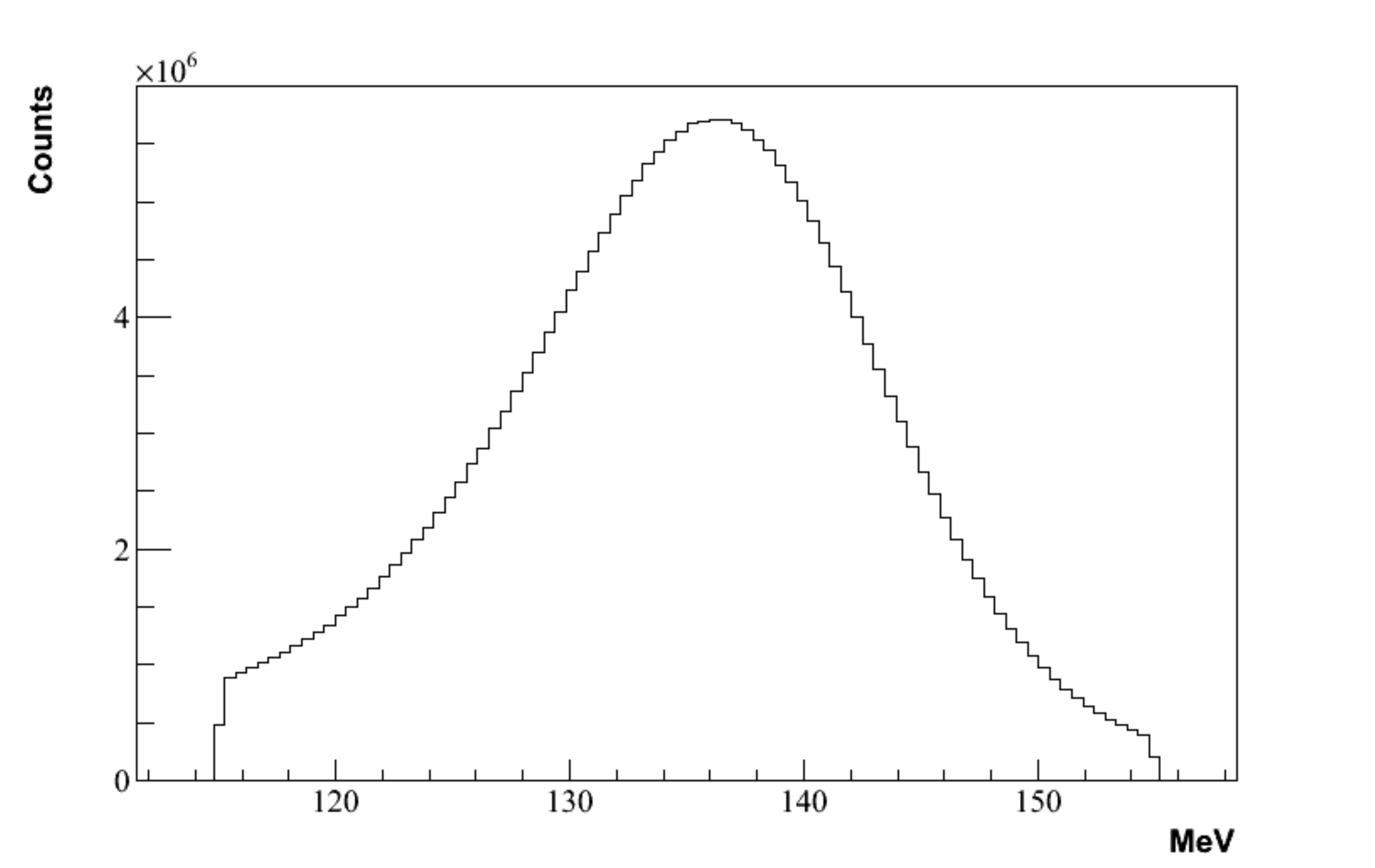}
	\end{center}
    \caption{Reconstruction of the $\pi^0$ invariant mass from its decay of two photons. }
      \end{figure}

  \section{Data analysis}
 Final state products from the reaction $\gamma p \rightarrow p + \pi^0 $ are initially identified in the data analysis. The $\pi^0$  can not be directly measured due to its short life time. The $\pi^0$  decays via the electromagnetic interaction, $\pi^0\rightarrow \gamma\gamma$. This allows the $\pi^0$ to be reconstructed from its decay products, detected in the CB, demonstrated in figure 2. With the information of the 4-vector incident photon, target and outgoing $\pi^0$ , the proton can be reconstructed by the missing mass technique, identifying the reaction to be studied. 
 
 \begin{equation}
p_{recon}=p_{tagged}+p_{target}-p_{meson}
\end{equation}

Where $p_{tagged}$, $p_{target}$ and $p_{meson}$ are the 4-vectors of the tagged photon from the beam, target and meson respectively. An example of the $\pi^0$ missing mass (proton) can be seen on the right of figure 3. The proton peak is not a clean sharp peak, due to the interactions of the photon with the carbon and oxygen atoms in the butanol target. To obtain a clean peak the background is studied via a separate data run on a carbon target. The same method was used to identify the $\pi^0$  and $\pi^0$  missing mass.  Since the beam time on the carbon target was less than the butanol, the carbon must be scaled the match the butanol background. The $\pi^0$ missing mass distribution from butanol and carbon were divided and the ratio in the bound nucleon region (at lower mass) is used to determine the scaling factor, seen in the left plot of figure 3. This factor is then used to scale the carbon data to the butanol, as seen in the right plot of figure 3. The $\pi^0$ missing mass from butanol after subtraction of the scaled carbon can been seen in figure 4. A gaussian is fitted to the final histogram, of the clean proton peak, and a $3\sigma$ cut can be made on the missing mass resulting in the final event selection from which F and T will be determined.
      \begin{figure}[!h]
      \begin{center}
      \includegraphics[height=.25\textheight]{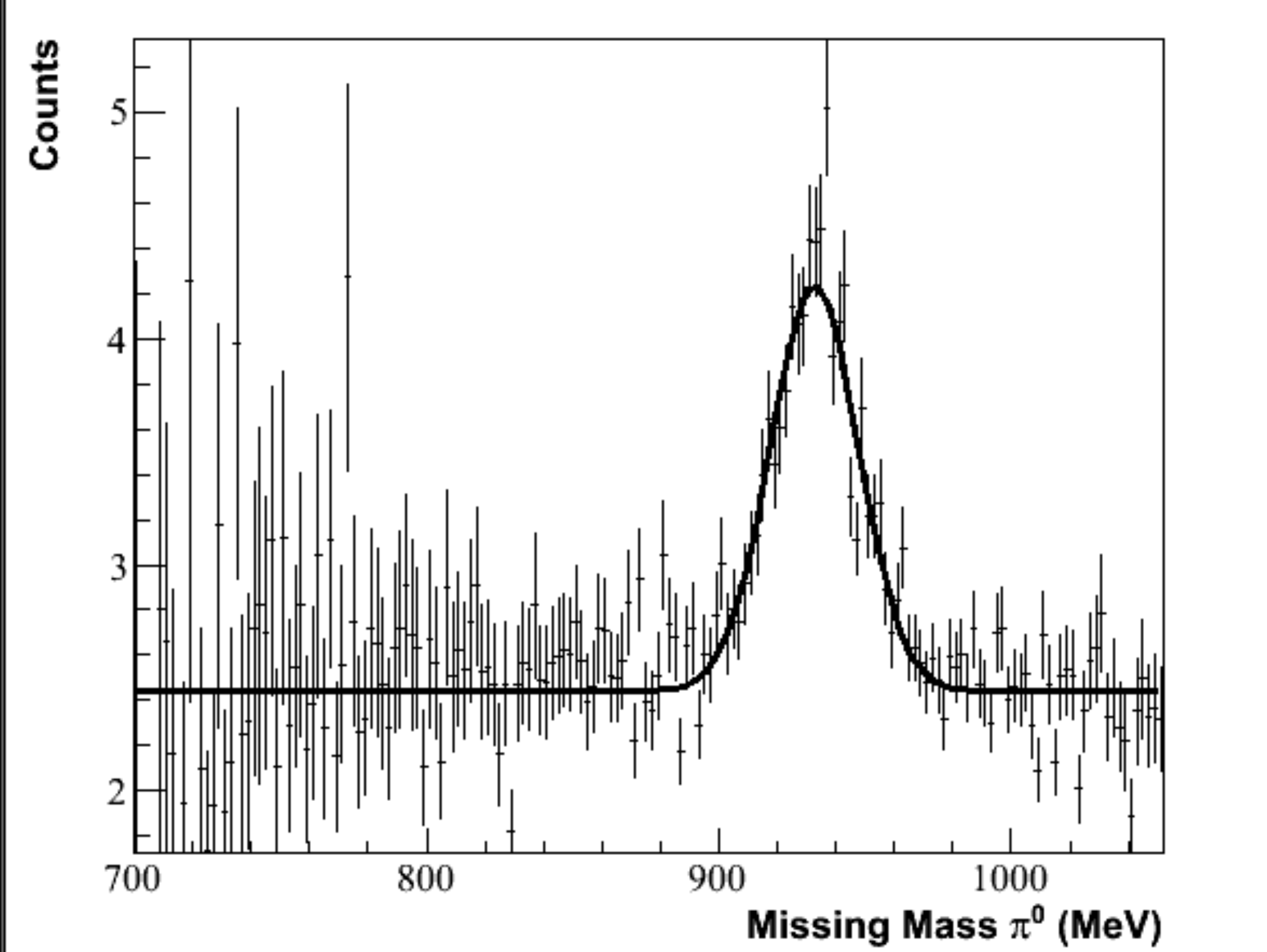}
\includegraphics[height=.25\textheight]{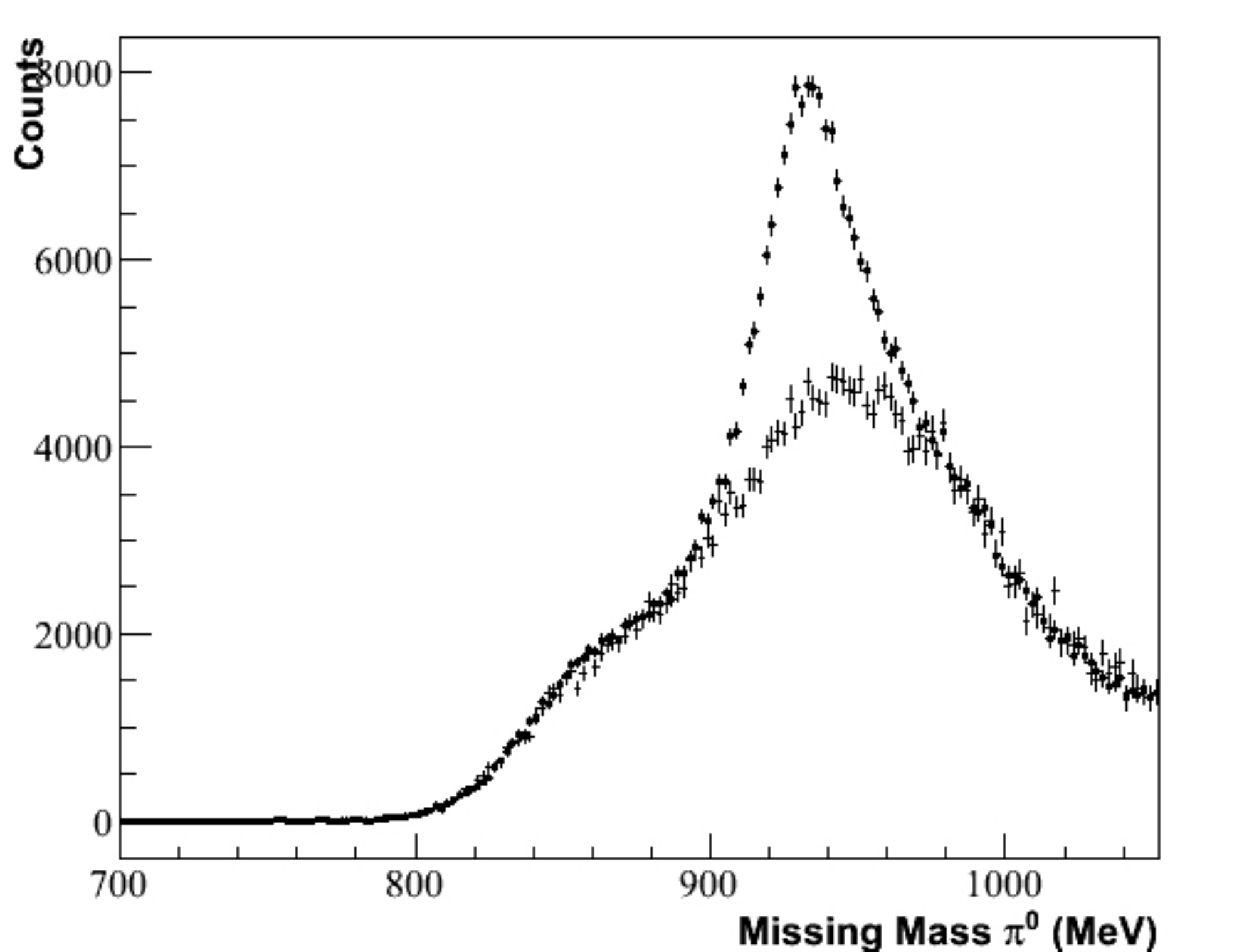}
      \end{center}
    \caption{The left hand plot  shows the two distributions are normalised via their ratio. The right plot demonstrates  the $\pi^0$ missing mass distributions (peaked on the proton mass) for butanol (black circles) and carbon (scaled by ratio technique) targets.}
      \end{figure}
     
          \begin{figure}[!h]
          \begin{center}
          \includegraphics[height=.25\textheight]{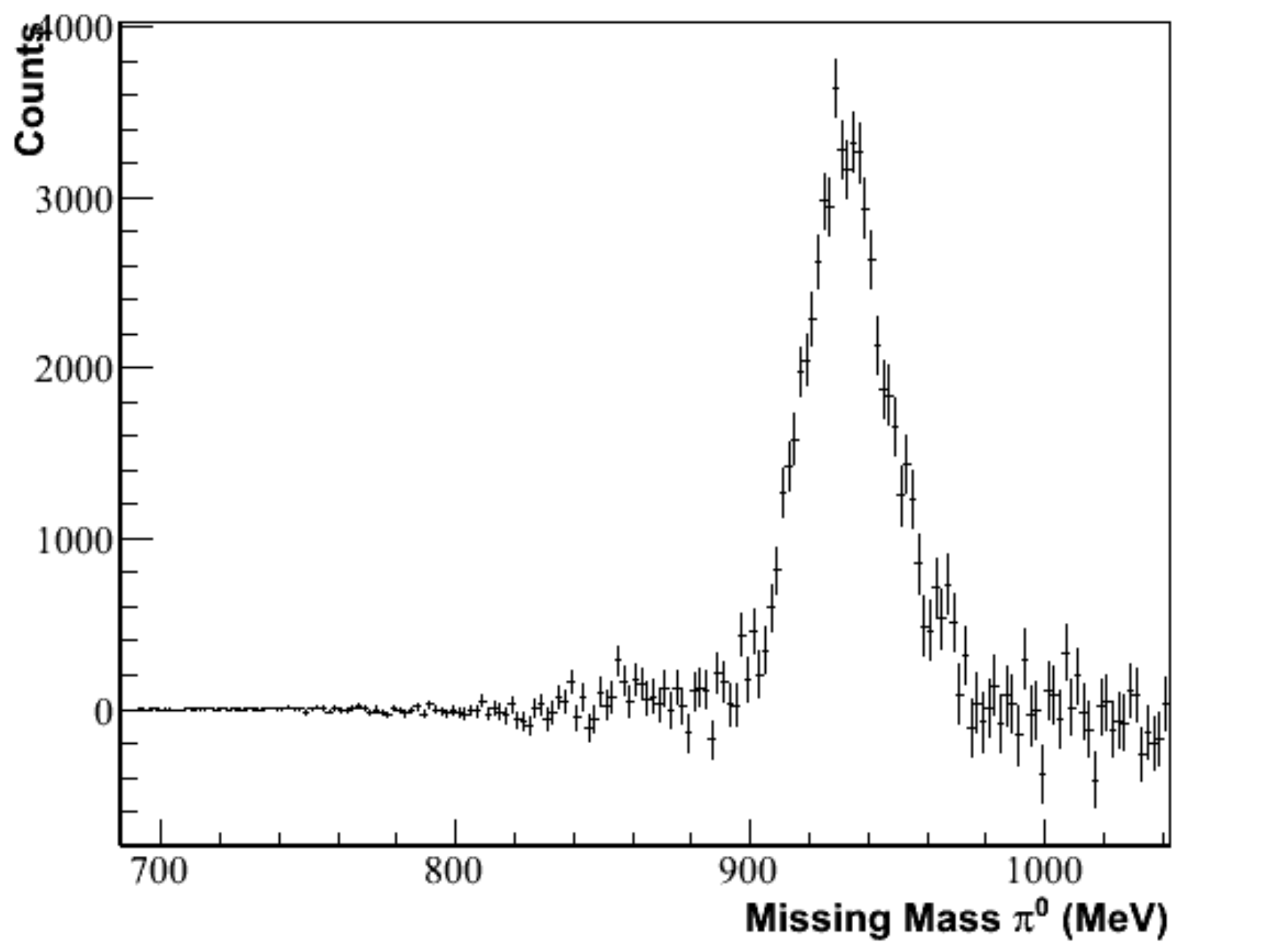}
          \end{center}
              \caption{   Resulting subtracted spectra, i.e. the free proton.}
      \end{figure}

 \section{Extraction of polarisation observables}
Once the  $\gamma p \rightarrow p + \pi^0 $ reaction was identified with selected cuts, the final step is to determine the $\pi^0$ azimuthal distribution for each beam and target polarisation state.  

With a transversely polarised target and circular polarised beam, the differential cross section is the following:

\begin{equation}
\frac{d\sigma}{d\Omega}=(\frac{d\sigma}{d\Omega})_{unpol}(1+P_xP_c cos(\phi_0-\phi)F+P_y sin(\phi_0-\phi)T)
\end{equation}

The two possible settings for the beam settings are, with a helicity of 1 and -1, for the target, pointing vertically up an down, known as positive and negative respectively. $P_{C}$ is the polarisation of the beam, $P_{x}$ and $P_{y}$ are the polarisation of the target, $\phi_0$ is the direction of the target and  F and T are the polarisation observables. Asymmetry relations can then be constructed and fitting functions used to extract the polarisation observables of interest with the following formulae:

\begin{equation}
Beam Asymmetry=\frac{\sigma^+ - \sigma^-}{\sigma^+ + \sigma^-} =\frac{P_TP_c cos(\phi_0-\phi)F}{1+P_Tsin(\phi_0-\phi)T}
\end{equation}

\begin{equation}
Target Asymmetry=\frac{\sigma^\uparrow - \sigma^\downarrow}{\sigma^\uparrow + \sigma^\downarrow} =P_T sin(\phi_0-\phi)T
\end{equation}

Where $\sigma^+$ and $\sigma^-$ are the $\phi$ distributions of the $\pi^0$ with +1 and -1 beam helicity, $\sigma^\uparrow$ and $\sigma^\downarrow$ are the $\phi$ distributions with target polarised up and down respectively. Figure 5 shows an example of beam and target asymmetries, fitted with the above equations to extract F and T. 

Since NMR calculates the polarisation of the hydrogen atoms, it does not take into account the unpolarised nuclei in the oxygen and carbon in the butanol. The quasi-free scattering from the protons of these nuclei contribute an unpolarised background to the measured asymmetry. Therefore, $P_{T}=p_{T}f$, is an effective polarisation, where f is the dilution factor.

The dilution factor, the number of polarised nuclei in the butanol target, is measured for each energy and $cos(\theta)$ bin. The carbon target was used to model the background, unpolarised nuclei (including oxygen). The integral of the butanol and scaled carbon spectrum, within proton cuts, allows the dilution factor to be determined by:

\begin{equation}
f(E_{\gamma},\theta)= \frac{N_{B}-N_{C}}{N_{C}}
\end{equation}
 Where $N_{B}$ and $N_{C}$ are the integrals within selected missing mass cuts for the butanol and carbon spectra directly. 

An example of preliminary results for F and T in the photon energy bin 350-375 MeV is shown in figure 6 along with a comparison to MAID. Further refinement of the analysis is underway with emphasis on finer energy binning and normalisation of the data.

    \begin{figure}[!h]
    \begin{center}
\includegraphics[height=.2\textheight]{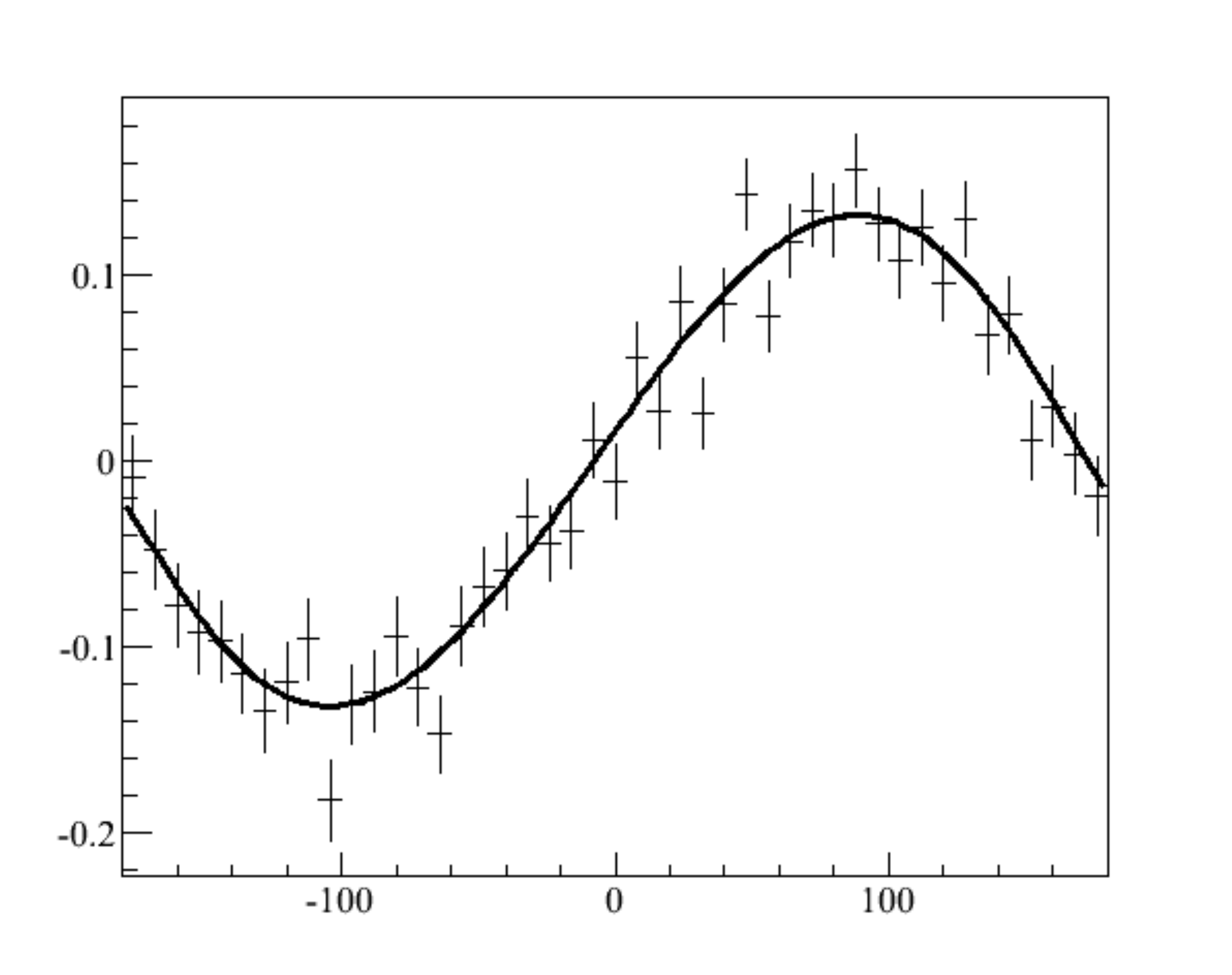}
          \includegraphics[height=.2\textheight]{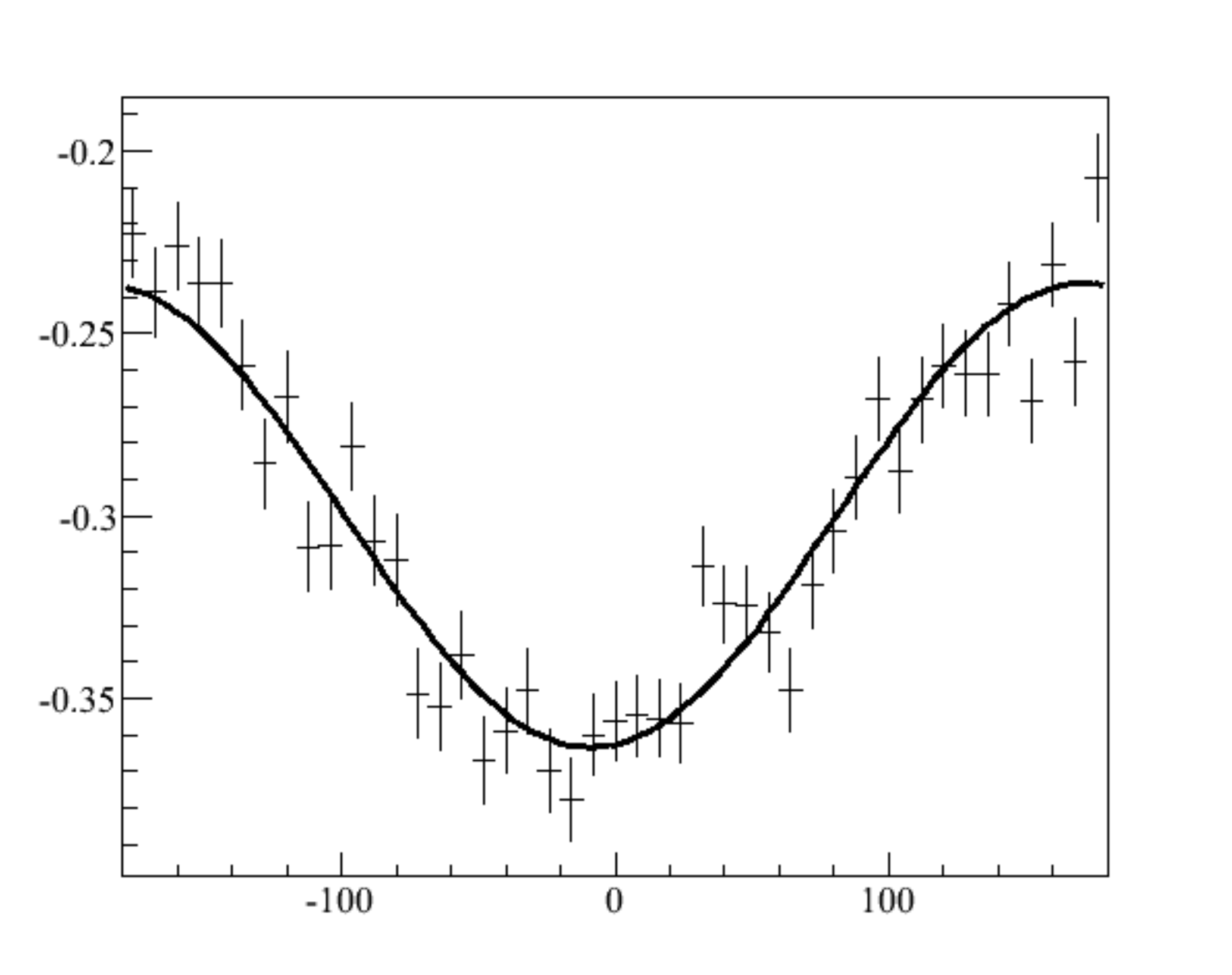}
          \end{center}
    \caption{350-375 MeV photon energy bin, $\theta_{CM}=57^0$ polar angle bin . Beam asymmetry (left hand plot) fitted with eqn. 3 and target asymmetry  (right hand plot) fitted with eqn. 4. }
      \end{figure}

    \begin{figure}[!h]
    \begin{center}
\includegraphics[height=.2\textheight]{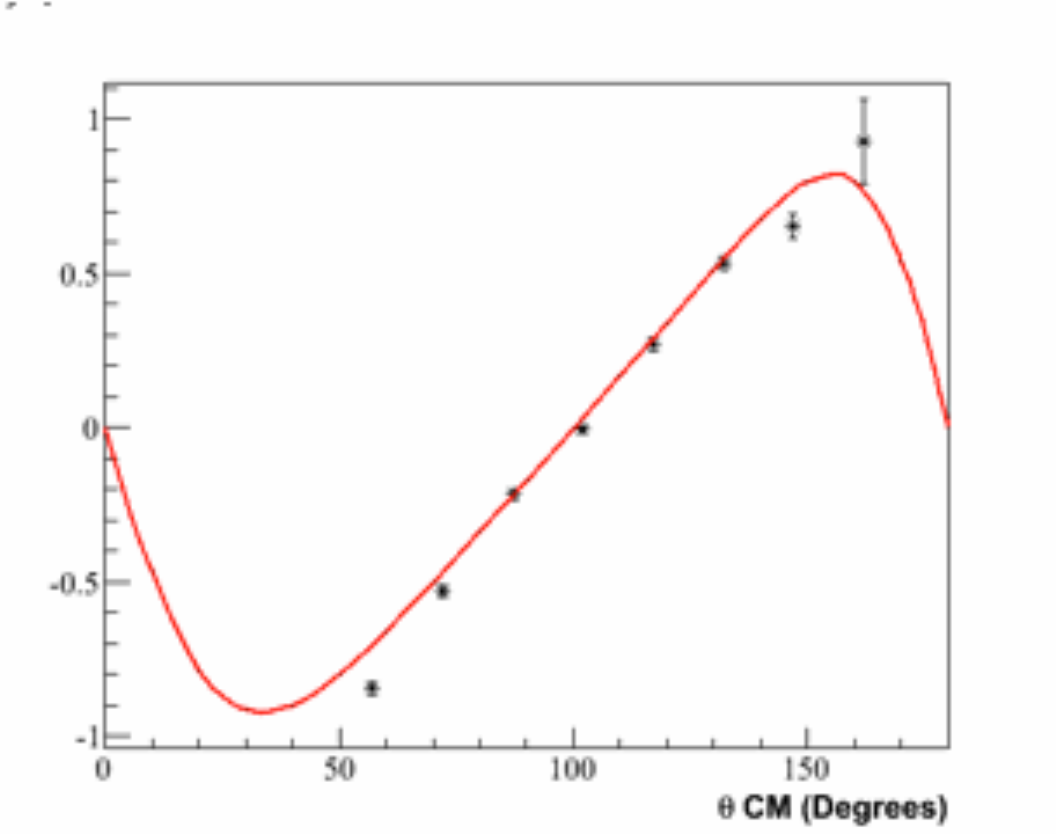}
          \includegraphics[height=.2\textheight]{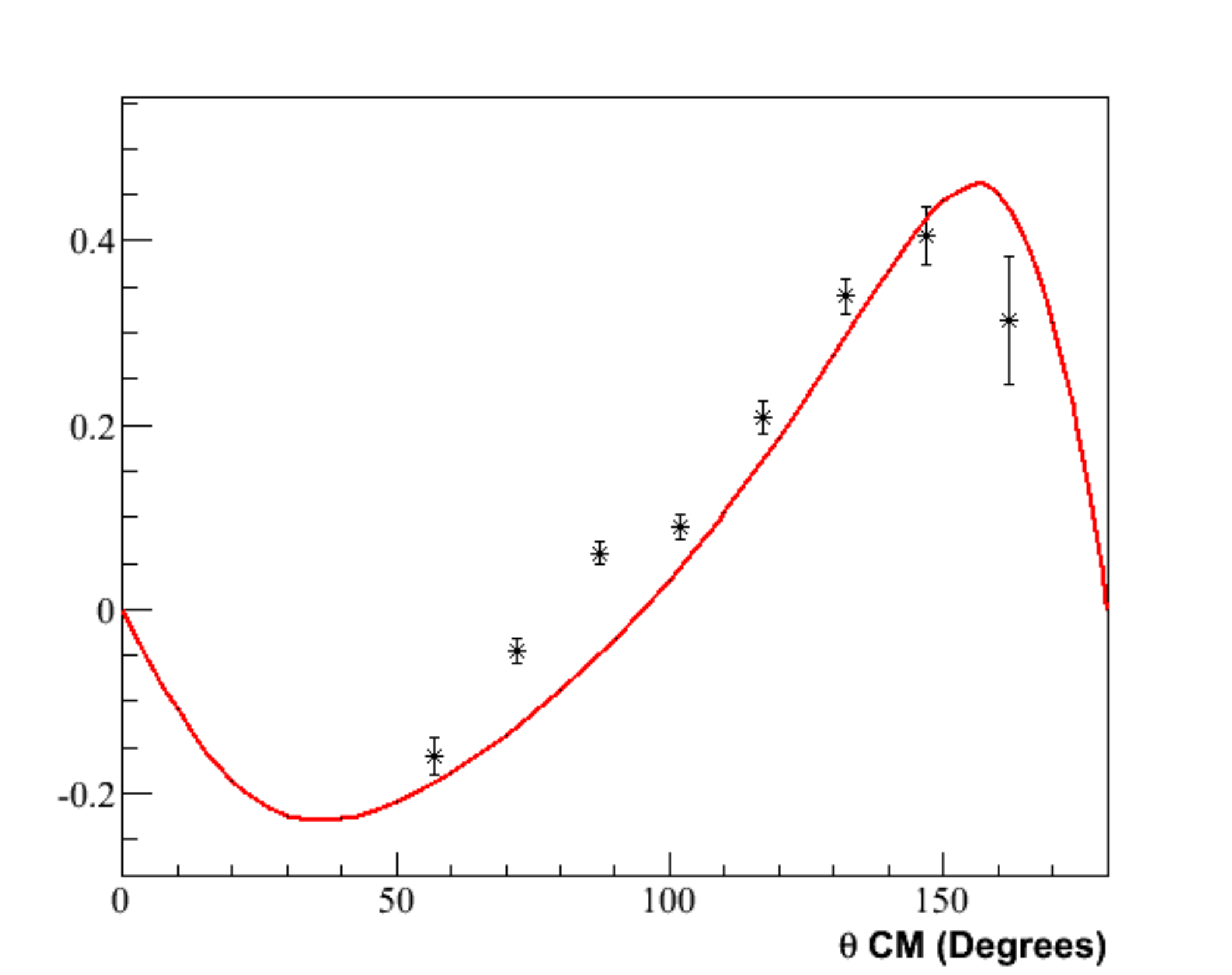}
          	\end{center}
    \caption{350-375 MeV photon energy bin. F (left hand plot) and T (right hand plot) - preliminary results. Curve is MAID}
      \end{figure}

The inclusion of this data will allow the further constraining of theoretical models and contribute the F and T observables up to 400 MeV beam energy to the global database. Moreover, comparison of the $\pi^0$ and $\pi^+$ observables will allow a search for isospin breaking effects in these observables. Analysis for the $n\pi^+$ channel will be carried out in parallel.

\section{Acknowledgements}
The author would like to acknowledge the support provided by members of the A2 collaboration at MAMI, staff at Edinburgh University's nuclear physics group and the UK Science and Technology Facilities Council.

\end{document}